\documentclass[showpacs,nofootinbib,showkeys,eqsecnum,prd,aps,preprint]{revtex4-1}
\usepackage{amssymb}
\usepackage{amsmath}
\usepackage{amsfonts}
\usepackage{graphicx}
\usepackage{bm}
\usepackage[T1]{fontenc}
\usepackage{graphicx}

\input{tcilatex}
\begin{document}

\title{\textbf{Particle creation from the vacuum by an exponentially
decreasing electric field}}
\author{T. C. Adorno${}^{d}$}
\email{tadornov@usp.br}
\author{S.~P.~Gavrilov${}^{a,c,d}$}
\email{gavrilovsergeyp@yahoo.com}
\author{D.~M.~Gitman${}^{a,b,d}$}
\email{gitman@if.usp.br}
\date{\today }

\begin{abstract}
We analyze the creation of fermions and bosons from the vacuum by the
exponentially decreasing in time electric field in detail. In our
calculations we use QED and follow in main the consideration of particle
creation effect in a homogeneous electric field. To this end we find
complete sets of exact solutions of the $d$-dimensional Dirac equation in
the exponentially decreasing electric field and use them to calculate all
the characteristics of the effect, in particular, the total number of
created particles and the probability of a vacuum to remain a vacuum. It
should be noted that the latter quantities were derived in the case under
consideration for the first time. All possible asymptotic regimes are
discussed in detail. In addition, switching on and switching off effects are
studied.
\end{abstract}

\maketitle

\affiliation{$^{a}$Department of Physics, Tomsk State University, 634050, Tomsk, Russia\\
$^{b}$P. N. Lebedev Physical Institute, 53 Leninskiy prospect, 119991,
Moscow, Russia\\
${}^{c}$Department of General and Experimental Physics, Herzen State
Pedagogical University of Russia, Moyka embankment 48, 191186
St.~Petersburg, Russia\\
$^{d}$Institute of Physics, University of S\~{a}o Paulo, CP 66318, CEP
05315-970 S\~{a}o Paulo, SP, Brazil;\\
}


\section{Introduction\label{S1}}

Particle creation from the vacuum by strong external electromagnetic fields
is an important nonperturbative effect, theoretical study of which has a
long history, see for example the Refs. ~\cite%
{Sch51,Nik70,Gitman77,GMR85,FGS,BirDav82,GavGT06}. To be observable, the
effect needs very strong electric fields in magnitudes compared with the
Schwinger critical field $E_{c}=m^{2}c^{3}/e\hbar \simeq 1.3\times 10^{16}\,%
\mathrm{V}\cdot \mathrm{cm}^{-1}$. However, recent progress in laser physics
allows one to hope that the nonperturbative regime of pair production may be
reached in the near future, see Ref.~\cite{Dun09} for the review.
Electron-hole pair creation from the vacuum becomes also an observable in
the laboratory effect in graphene physics, an area that is currently under
intense development \cite{castroneto,dassarma}. In particular, this effect
is crucial for understanding the conductivity of the graphene, especially in
the so-called nonlinear regime, see, for example, Ref.~\cite{GavGitY12}. The
particle creation from the vacuum by external electric and gravitational
backgrounds plays also an important role in cosmology and astrophysics~\cite%
{BirDav82}.

It should be noted that the particle creation from the vacuum by external
fields is a nonperturbative effect and its calculation essentially depends
on the structure of the external fields. Sometimes calculations can be done
in the framework of the relativistic quantum mechanics, sometimes using
semiclassical and numerical methods (see Refs. \cite{BirDav82,Dun09,DumDun13}
for the review). The vast majority of analytic works in this area, in QED,
is based on the worldline and instanton formalisms, rather than solving the
Dirac equation (for example, see \cite{DS05,I14} and references therein). In
all these cases the authors calculate, in fact, the one-loop effective
action, whose imaginary part is related to the probability of a vacuum to
remain a vacuum. However, in those cases, when the semiclassical
approximation does not work,{\Large \ }the most convinced consideration is
formulated in the framework of QFT, in particular, in the framework of QED,
see Ref. \cite{Gitman77,FGS,GavGT06}. In particular, in the latter approach
nonperturbative calculations are based on the existence of exact solutions
of the Dirac equation with the corresponding external electromagnetic field.
In fact, until now, there are known\emph{\ }only few exactly solvable cases
for either time-dependent homogeneous or constant inhomogeneous electric
fields. One of them is related to the constant uniform electric field \cite%
{Sch51}, another one to\ the\ so-called adiabatic electric field $E\left(
t\right) =E\cosh ^{-2}\left( t/\alpha \right) \,$\cite{NarNik70} (see also 
\cite{DunHal98}), the case related to the so-called $T$-constant electric
field \cite{BagGiS75,GavGit96,GavGit08}, which corresponds to a constant
electric field that turns-on and -off at definite times instants $t_{1}$ and 
$t_{2}$,\ ($t_{2}-t_{1}=T$) being constant inside of the time interval $T$,
the case related to a periodic alternating electric field \cite{NarNIk74},
and the number of a constant inhomogeneous electric fields of the similar
forms where time $t$ is replaced by the spatial coordinate $x$. To complete
the picture, we note that these exist exact solutions of the Dirac equation
with some electric fields satisfying more complicated symmetries, e.g. with
potentials given in the light-cone variables, for example, see \cite{TTW01}
and \cite{BagGi90}.{\Large \ }The existence of exactly solvable cases of
particle creation is extremely important both for deep understanding of QFT
in general and for studying quantum vacuum effects in the corresponding
external fields.

In this article, we present a new exactly solvable case of particle creation
that corresponds to the so-called $T$-exponentially decreasing in time
electric field, which switches on at the time instant $t_{1}$, switches off
at the time instant $t_{2}$ ($t_{2}-t_{1}=T$), and within the time interval $%
T$ has the form $E_{x}\left( t\right) =Ee^{-k_{0}\left( t-t_{1}\right) }$,
where $k_{0}$, $E$ are some positive constants. In particular, this field
presents the example of an exponentially decaying electric field when $%
t_{2}\rightarrow \infty $. Technically this exactly solvable case differs
essentially from all the above mentioned cases because of an asymmetrical
asymptotic behavior of the external electric field. Consideration of such a
case has an interesting physical motivation. The corresponding external
electric field can be treated as one, which is created by an external
current that switches on fast enough and then is slowly switching off
(decreases) because of some dissipation processes. One can demonstrate that
under certain conditions the main contribution to particle creation is due
to the decreasing part of the electric field, whereas the contribution from
the increasing part of the field is relatively small. The qualitative
difference in the asymptotic behavior of the external electric field under
consideration allows one to study the role of switching on and switching off
for an electric field\emph{.} We just from the beginning consider general $%
(d=D+1)$-dimensional Minkowski space-time, to be able to use the case $D=3$\
for describing high-energy effects, while the case $D=1,2,3$\ could be
adequate for condense matter problems. For completeness, the case of scalar
particles is considered too.

It is worth to note that the differential mean number of particles created
by a kind of exponentially decaying electric field was calculated previously%
{\Large \ }in the Ref.{\Large \ }\cite{Spokoinyi} in the framework of some
semiclassical considerations and in \cite{BRW11} using the
Dirac-Heisenberg-Wigner function.{\Large \ }However, the authors of the
latter work did not present any analysis how their results depend on the
problem parameters in the case of a strong field, in fact, they studied the
weak field limit only.

As was already said, in our calculations, we use the general theory of Ref. 
\cite{Gitman77,FGS} and follow in main the consideration of particle
creation effect in a homogeneous electric field \cite{GavGit96}, see
appendix \ref{App} for some basic elements. To this end we find complete
sets of exact solutions of the Dirac and Klein-Gordon equations in the $T$%
-exponentially decreasing electric field and use them to calculate all the
characteristics of the effect, in particular, the differential mean number
of particle created, total number of created particles, and the probability
for a vacuum to remain a vacuum. It should be noted that the latter
quantities were derived in the case under consideration for the first time.
Using these solutions, we analyze particle creation in the case of the
exponentially decaying electric field. All possible asymptotic regimes are
discussed in detail. In addition, switching on and switching off effects are
studied.

\section{Exponentially decreasing electric field\label{S2}}

We consider the Dirac equation\footnote{%
From this section and in what follows we are considering the system of units 
$\hslash =c=1$ and the fine structure constant is $\alpha =e^{2}$.} in $%
(d=D+1)$-dimensional Minkowski space with an external electromagnetic field
given by potentials $A_{\mu }\left( x\right) $,%
\begin{equation}
\left( \gamma ^{\mu }\hat{P}_{\mu }-m\right) \psi \left( x\right) =0\,,\ \ 
\hat{P}_{\mu }=\hat{p}_{\mu }-qA_{\mu }\left( x\right) ,\ \ \hat{p}_{\mu
}=i\partial _{\mu }\,.  \label{iint0}
\end{equation}%
Here $\psi (x)$ is a $2^{[d/2]}$-component spinor ($[d/2]$ stands for the
integer part of $d/2$), $m$ is the particle mass, $q$ is the particle charge
(for the electron $q=-e,$ with $e>0$ being the absolute value of the
electron charge), $x=(x^{\mu })=(x^{0},\mathbf{x}),\;\mathbf{x}=(x^{i}),$ $%
x^{0}=t,$ the Greek and Latin indexes assume values $\mu =0,1,\dots ,D$ and $%
i=1,\ldots ,D$ respectively, and $\gamma $-matrices satisfy the standard
anticommutation relations: 
\begin{equation*}
\left[ \gamma ^{\mu },\gamma ^{\nu }\right] _{+}=2\eta ^{\mu \nu },\ \ \eta
_{\mu \nu }=\mathrm{diag}(1,-1,\ldots ,-1).
\end{equation*}

Using the Ansatz $\psi \left( x\right) =\left( \gamma ^{\mu }\hat{P}_{\mu
}+m\right) \phi \left( x\right) $, one finds that the spinor $\phi \left(
x\right) $ satisfies the following equation:%
\begin{align}
& \left( \hat{P}^{2}-m^{2}-\frac{q}{2}\sigma ^{\mu \nu }F_{\mu \nu }\right)
\phi \left( x\right) =0\,,  \notag \\
& \sigma ^{\mu \nu }=\frac{i}{2}\left[ \gamma ^{\mu },\gamma ^{\nu }\right]
\,,\ \ F_{\mu \nu }=\partial _{\mu }A_{\nu }-\partial _{\nu }A_{\mu }\,.
\label{int3}
\end{align}

In what follows, we consider the so-called $T-$exponentially decreasing
electric field with a constant direction along the $x$ axis. This field
switches on at $t_{1}$ and switches off at $t_{2},$ being nonzero within the
time interval $T=t_{2}-t_{1}>0$ and zero outside of it,%
\begin{equation}
E_{x}\left( t\right) =E\left\{ 
\begin{array}{l}
0\,,\ \ \ \ \ \ \ \ \ \ \ t\in \mathrm{I}=\left( -\infty ,t_{1}\right) \\ 
e^{-k_{0}\left( t-t_{1}\right) }\,,\ \ t\in \mathrm{II}=\left[ t_{1},t_{2}%
\right] \\ 
0\,,\ \ \ \ \ \ \ \ \ \ \ t\in \mathrm{III}=\left( t_{2},+\infty \right)%
\end{array}%
\right. ,\ \ k_{0}>0.  \label{expfield}
\end{equation}%
We choose the corresponding potentials as $A^{\mu }(t)=\delta _{1}^{\mu
}A_{x}(t)$ with only one nonzero component,%
\begin{equation}
A_{x}\left( t\right) =\frac{E}{k_{0}}\left\{ 
\begin{array}{l}
1,\ \ \ \ t\in \mathrm{I} \\ 
e^{-k_{0}\left( t-t_{1}\right) }\,,\ \ t\in \mathrm{II} \\ 
e^{-k_{0}T}\,,\ \ \ \ t\in \mathrm{III}%
\end{array}%
\right. .\,  \label{v1}
\end{equation}%
We admit that the switching off can occur in the remote future such that $%
t_{2}$ can be infinite, $t_{2}=+\infty $, under the condition that $t_{1}$\
remains finite.

\begin{figure}[th]
\includegraphics[scale=0.5]{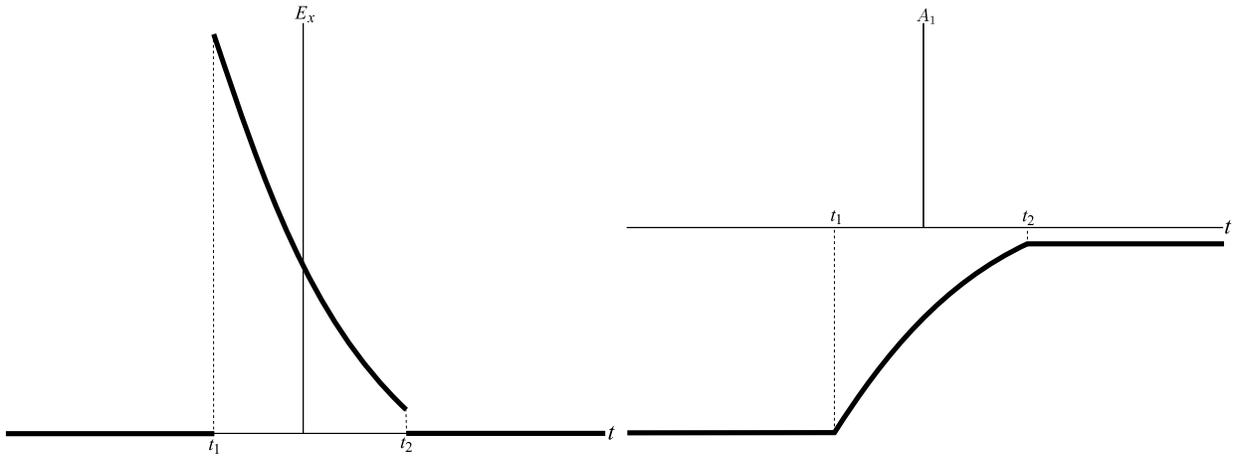}
\caption{The exponentially decreasing electric field and its potential.}
\end{figure}

Solving equation (\ref{int3}), we will use a set of constant orthonormalized
spinors $v_{s,\sigma }$,%
\begin{equation*}
v_{s,\sigma }^{\dagger }v_{s^{\prime },\sigma ^{\prime }}=\delta
_{s,s^{\prime }}\delta _{\sigma ,\sigma ^{\prime }},\ \ vv^{\dagger }=%
\mathbb{I},
\end{equation*}%
with $s=\pm 1$, and $\sigma =(\sigma _{1},\sigma _{2},\dots ,\sigma
_{\lbrack d/2]-1})$, $\sigma _{j}=\pm 1$, such that $\gamma ^{0}\gamma
^{1}v_{s,\sigma }=sv_{s,\sigma }$. For the $d\geq 4$ the indices $\sigma
_{j} $ describe the spin polarization, which is not coupled to the electric
field, and together with the additional index $s$ provide a suitable
parametrization of the solutions. Note that for the $d=2,3$ there is only
one spin degree of freedom and the spinors are labeled either by $s=+1$ or
by $s=-1$. Solutions of eq. (\ref{int3}) with the potential given by eq. (%
\ref{v1})\ can be found in the form%
\begin{equation}
\phi _{\mathbf{p,}s,\sigma }(x)=\varphi _{\mathbf{p,}s}(t)e^{i\mathbf{p}%
\cdot \mathbf{x}}v_{s,\sigma }\,,  \label{int4}
\end{equation}%
where scalar functions $\varphi _{\mathbf{p,}s}\left( t\right) $ satisfy the
following second order differential equation%
\begin{equation}
\left[ \frac{d^{2}}{dt^{2}}+\left[ p_{x}-qA_{x}\left( t\right) \right] ^{2}+%
\mathbf{p}_{\perp }^{2}+m^{2}+isqE_{x}\left( t\right) \right] \varphi _{%
\mathbf{p,}s}(t)=0\,,  \label{t3}
\end{equation}%
where $\mathbf{p}_{\perp }$ is the transversal particle momentum, $\mathbf{p}%
_{\perp }=(0,p^{2},\dots ,p^{D})$.

Thus, in what follows, we are going to deal with two complete set of
solutions of the Dirac equation (\ref{iint0}) of the following structure%
\begin{eqnarray}
\left. _{\pm }\!\psi _{n}\left( x\right) \right. &=&\left( \gamma ^{\mu }%
\hat{P}_{\mu }+m\right) \left. _{\pm }{}\phi _{\mathbf{p},s,\sigma }\left(
x\right) \right. \,,  \notag \\
\left. ^{\pm }\!\psi _{n}\left( x\right) \right. &=&\left( \gamma ^{\mu }%
\hat{P}_{\mu }+m\right) \left. ^{\pm }{}\phi _{\mathbf{p},s,\sigma }\left(
x\right) \right. \,,  \label{t4.10}
\end{eqnarray}%
where spinors $_{\pm }{}\phi _{\mathbf{p},s,\sigma }\left( x\right) $ and $%
^{\pm }{}\phi _{\mathbf{p},s,\sigma }\left( x\right) $ are given by Eq.~(\ref%
{int4}) with solutions $_{\zeta }\varphi _{\mathbf{p},s}(t)$ and $^{\zeta
}\varphi _{\mathbf{p},s}(t)$, respectively, satisfying Eq.~(\ref{t3}) with
initial or final conditions that are specified in which follows. Here we
denote by $n=\left( \mathbf{p},\sigma \right) $ a complete set of quantum
numbers of the Dirac spinor for both cases $s=+1$\ or\emph{\ }$s=-1$\emph{. }
Note that for the $d\geq 4$ the Dirac spinors given by the choice of $s=+1$
in (\ref{t4.10}) are linearly dependent with the spinors given by the choice
of $s=-1$. Thus, one can form physically equivalent complete sets of the
Dirac spinors for both choices of parametrization.{\Huge \ }The algebra of $%
\gamma $-matrices has two inequivalent representations in $d=3$ dimensions,
representation given by $s=+1$ and $s=-1$ are associated with different
fermion species.

Note that a formal reduction to the spinless case that corresponds to the
use of the Klein-Gordon equation instead of the Dirac one can be done by
setting $s=0$\ in (\ref{t3}) and $v_{s,\{r\}}=1$\ in (\ref{int4}). In this
case $n=(\mathbf{p)}$.

In the first region $\mathrm{I}$ and in the third region $\mathrm{III}$ the
electric field is absent and Eq.~(\ref{t3}) has plane wave solutions $%
_{\zeta }\varphi _{\mathbf{p},s}(t)$ and $\ ^{\zeta }\varphi _{\mathbf{p}%
,s}(t)$, respectively, with additional quantum number $\zeta =\pm $, which
satisfy simple dispersion relations%
\begin{align}
& \mathrm{I}:\ _{\zeta }\varphi _{\mathbf{p},s}(t)\sim e^{-i\zeta
p_{0}\left( t_{1}\right) t}\,,\quad \mathrm{III}:\ ^{\zeta }\varphi _{%
\mathbf{p},s}(t)\sim e^{-i\zeta p_{0}\left( t_{2}\right) t}\,,  \notag \\
& p_{0}\left( t\right) =\sqrt{\left( p_{x}^{\prime }-\frac{\left\vert
qE\right\vert }{k_{0}}e^{-k_{0}\left( t-t_{1}\right) }\right) ^{2}+\mathbf{p}%
_{\perp }^{2}+m^{2}}\,,  \label{free-solutions}
\end{align}%
where $p_{x}^{\prime }=\varkappa p_{x}$\ and\ $\varkappa =\mathrm{sgn}\left(
qE\right) $. Here, the quantum numbers $\zeta $ label particle/antiparticles
states such that positive $\left( \zeta =+\right) $/negative $\left( \zeta
=-\right) $ values define particles/antiparticles states, respectively.

In the second region $\mathrm{II}$, it is convenient to introduce a new
variable $\eta $ and represent the functions $\varphi _{\mathbf{p,}s}$\ as%
\begin{align}
\eta & =\frac{2i\left\vert qE\right\vert }{k_{0}^{2}}e^{-k_{0}\left(
t-t_{1}\right) },\ \ \varphi _{\mathbf{p,}s}\left( t\right) =e^{-\eta
/2}\eta ^{\nu }\chi _{\mathbf{p,}s}\left( \eta \right) ,  \notag \\
\nu & =i\frac{\omega _{0}}{k_{0}}\,,\ \ \omega _{0}=\sqrt{\mathbf{p}%
^{2}+m^{2}}.  \label{v3}
\end{align}%
Then the functions $\chi _{\mathbf{p,}s}\left( \eta \right) $ satisfy the
confluent hypergeometric equation \cite{Ederlyivol2},%
\begin{equation*}
\left[ \eta \frac{d^{2}}{d\eta ^{2}}+\left( c-\eta \right) \frac{d}{d\eta }-a%
\right] \chi _{\mathbf{p,}s}\left( \eta \right) =0\,,
\end{equation*}%
with parameters 
\begin{equation}
c=1+2\nu \,,\ \ a=\frac{1}{2}\left( 1-\varkappa s\right) -\frac{%
ip_{x}^{\prime }}{k_{0}}+\nu \,.  \label{v5}
\end{equation}%
The complete set of solutions for this equation is formed by two linearly
independent confluent hypergeometric functions:%
\begin{equation*}
\Phi \left( a,c;\eta \right) \ \ \mathrm{and}\ \ \eta ^{1-c}\Phi \left(
a-c+1,2-c;\eta \right) ,
\end{equation*}%
where%
\begin{equation*}
\Phi \left( a,c;\eta \right) =1+\frac{a}{c}\frac{\eta }{1!}+\frac{a\left(
a+1\right) }{c\left( c+1\right) }\frac{\eta ^{2}}{2!}+\ldots \ \ .
\end{equation*}%
Thus one can find the general solution of the equation (\ref{t3}) in the
time region $\mathrm{II}$ as the following linear superposition%
\begin{align}
& \varphi \left( t\right) =a_{2}\varphi _{1}\left( t\right) +a_{1}\varphi
_{2}\left( t\right) ,  \notag \\
& \varphi _{1}\left( t\right) =e^{-\eta /2}\eta ^{\nu }\Phi \left( a,c;\eta
\right) ,\   \notag \\
& \varphi _{2}\left( t\right) =e^{-\eta /2}\eta ^{-\nu }\Phi \left(
a-c+1,2-c;\eta \right) ,\   \label{v6}
\end{align}%
where the constants $a_{1}$ and $a_{2}$ are fixed by initial conditions.

Taking into account expressions (\ref{free-solutions}) and (\ref{v6}), one
can construct orthonormalized solutions for the complete time interval in
the following form%
\begin{equation}
\left. ^{\zeta }\!\varphi _{\mathbf{p},s}\left( t\right) \right. =\left\{ 
\begin{array}{l}
g\left( _{-}|^{\zeta }\right) \ \left. _{-}C\right. e^{ip_{0}\left(
t_{1}\right) \left( t-t_{1}\right) }+g\left( _{+}|^{\zeta }\right) \ \left.
_{+}C\right. e^{-ip_{0}\left( t_{1}\right) \left( t-t_{1}\right) }\,,\ \
t\in \mathrm{I} \\ 
\,a_{2}^{\zeta }\varphi _{1}\left( t\right) +a_{1}^{\zeta }\varphi
_{2}\left( t\right) ,\ \ t\in \mathrm{II} \\ 
\left. ^{\zeta }C\right. e^{-i\zeta p_{0}\left( t_{2}\right) \left(
t-t_{2}\right) }\,,\ \ t\in \mathrm{III\,}%
\end{array}%
\right. ,  \label{v8.2}
\end{equation}%
where the constants $^{\zeta }C$ and $_{\zeta }C$ are defined by
normalization conditions for the Dirac spinors (\ref{t4.34}),

\begin{align}
& _{\zeta }C=\left[ 2Vp_{0}\left( t_{1}\right) p_{\zeta }\left( t_{1}\right) %
\right] ^{-1/2},\ \ ^{\zeta }C=\left[ 2Vp_{0}\left( t_{2}\right) p_{\zeta
}\left( t_{2}\right) \right] ^{-1/2},  \notag \\
& p_{\zeta }\left( t\right) =p_{0}\left( t\right) -\zeta s\left[
p_{x}-qA_{x}\left( t\right) \right] ,  \label{v10}
\end{align}%
and $p_{0}\left( t\right) $ is given by Eq.~(\ref{free-solutions}). Note
that notation $g\left( _{\zeta `}|^{\zeta }\right) $ corresponds to
definition (\ref{3.13}) from the Appendix. Coefficients $a_{1}^{\zeta }$, $%
a_{2}^{\zeta }$, $g\left( _{-}|^{\zeta }\right) $, and $g\left( _{+}|^{\zeta
}\right) $ are specified by the following gluing conditions: 
\begin{equation}
^{+}\varphi _{\mathbf{p},s}(t_{k}-0)=\;^{+}\varphi _{\mathbf{p}%
,s}(t_{k}+0),\quad \left. \partial _{t}\ ^{+}\varphi _{\mathbf{p},s}\left(
t\right) \right\vert _{t=t_{k}-0}=\left. \partial _{t}\ ^{+}\varphi _{%
\mathbf{p},s}\left( t\right) \right\vert _{t=t_{k}+0},\quad k=1,2.
\label{gl_cond}
\end{equation}

It can be seen from the consideration given in the appendix \ref{App} that
the probability of a vacuum to remain a vacuum, the probability of a
particle scattering, a pair creation, and a pair annihilation can be
expressed via the differential mean number of particles created from vacuum $%
N_{n}$ given by Eq.(\ref{difnumber}). From which it follows that one can
describe a vacuum instability for the case under consideration using the
quantity%
\begin{equation}
N_{\mathbf{p},\sigma }=\left\vert g(_{-}|^{+})\right\vert ^{2}  \label{difN}
\end{equation}%
only. Then it is enough to consider only the case $\zeta =+$ in Eq.(\ref%
{v8.2}). Using conditions (\ref{gl_cond}), we obtain%
\begin{equation*}
a_{1}^{+}=-\frac{i\left. ^{+}\!C\right. p_{0}\left( t_{2}\right) }{W}%
f_{1}\left( t_{2}\right) ,\ a_{2}^{+}=\frac{i\left. ^{+}\!C\right.
p_{0}\left( t_{2}\right) }{W}f_{2}\left( t_{2}\right) ,
\end{equation*}%
where $W$ is the corresponding Wronskian of the solutions \cite{Ederlyivol2},%
\begin{equation*}
W=\varphi _{1}\left( t\right) \frac{d}{dt}\varphi _{2}\left( t\right)
-\varphi _{2}\left( t\right) \frac{d}{dt}\varphi _{1}\left( t\right)
=2i\omega _{0},
\end{equation*}%
and%
\begin{equation}
f_{1,2}\left( t\right) =\left[ 1+\frac{ik_{0}\eta }{p_{0}\left( t\right) }%
\frac{d}{d\eta }\right] \varphi _{1,2}\left( t\right) .  \label{v8}
\end{equation}%
We finally find that the coefficient $g\left( _{-}|^{+}\right) $ takes the
form%
\begin{equation}
g\left( _{-}|^{+}\right) =\frac{1}{4\omega _{0}}\sqrt{\frac{p_{0}\left(
t_{2}\right) p_{-}\left( t_{1}\right) p_{0}\left( t_{1}\right) }{p_{+}\left(
t_{2}\right) }}\left[ f_{1}\left( t_{1}\right) f_{2}\left( t_{2}\right)
-f_{2}\left( t_{1}\right) f_{1}\left( t_{2}\right) \right] \,.  \label{v13}
\end{equation}%
One can demonstrate that in the case of a sufficient long duration of the
exponential electric field, when $t_{2}\rightarrow +\infty $ \ and $%
p_{0}\left( t_{1}\right) \left( t_{2}-t_{1}\right) \gg 1$, the differential
mean numbers, given by expression (\ref{v13}), coincide in the leading order
term approximation with the result obtained in the Ref. \cite{BRW11}.

Taking into account that the normalization constants $^{\zeta }C$\ and $%
_{\zeta }C$\ for the scalar case are%
\begin{equation*}
_{\zeta }\!C=\left[ 2Vp_{0}\left( t_{1}\right) \right] ^{-1/2},\ \ ^{\zeta
}\!C=\left[ 2Vp_{0}\left( t_{2}\right) \right] ^{-1/2},
\end{equation*}%
we find that in this case the coefficient $g\left( _{-}|^{+}\right) $\ has
the following form%
\begin{equation}
g\left( _{-}|^{+}\right) =\frac{1}{4\omega _{0}}\sqrt{p_{0}\left(
t_{2}\right) p_{0}\left( t_{1}\right) }\left[ f_{1}\left( t_{1}\right)
f_{2}\left( t_{2}\right) -f_{2}\left( t_{1}\right) f_{1}\left( t_{2}\right) %
\right] \,,  \label{v13b}
\end{equation}%
where $f_{1,2}\left( t\right) $\ are given by Eq.~(\ref{v8}) at $s=0$. The
differential mean number of created scalar particles is expressed via $%
g\left( _{-}|^{+}\right) $ (\ref{v13b}) as $N_{\mathbf{p}}=\left\vert
g(_{-}|^{+})\right\vert ^{2}$.

Expression (\ref{v13}) does not depend on spin polarization parameters $%
\sigma _{j}$. That is why all the probabilities and the mean number do not
depend on $\sigma _{j}$, so that the total (summed over all $\sigma _{j}$)
probabilities and the mean number are $J_{(d)}$ times greater than the
corresponding differential quantities. Here $J_{(d)}=2^{[\frac{d}{2}]-1}$ is
the number of spin degree of freedom. For example, the total number of
particles created with a given momentum $\mathbf{p}$ is 
\begin{equation}
\sum_{\sigma }N_{\mathbf{p},\sigma }=J_{(d)}\left\vert
g(_{-}|^{+})\right\vert ^{2}.  \label{e1a}
\end{equation}%
To get the total number $N$ of created particles one has to sum over the
spin projections, using eq.(\ref{e1a}), and then over the momenta. The
latter sum can be easily transformed into an integral, 
\begin{equation}
N=\sum_{\mathbf{p}}\sum_{\sigma }N_{\mathbf{p},\sigma }=\frac{VJ_{(d)}}{%
(2\pi )^{d-1}}\int d\mathbf{p}\left\vert g(_{-}|^{+})\right\vert ^{2},
\label{e1b}
\end{equation}%
where $V$ is ($d-1$)-dimensional spatial volume.

The expression above depends essentially on the time interval of the field
duration $T=t_{2}-t_{1}$. Then the effect of pair creation depends on two
dimensionless parameters $k_{0}T$ and $\left\vert qE\right\vert /k_{0}^{2}$.
For $k_{0}$ fixed, the first allows one to analyze the characteristics with
respect to the time duration $T$\ of the electric field, while the second
provides information on the maximum magnitude of the field $\left\vert
E\right\vert $, for $k_{0}$ fixed too.

\section{Exponentially decaying strong field\label{S3}}

\subsubsection{Differential quantities}

Let us consider the exponentially decaying electric field given by Eq.~(\ref%
{v1}), with%
\begin{equation}
\frac{\left\vert qE\right\vert }{k_{0}^{2}}e^{-k_{0}T}\,\ll 1\,,
\label{v14.1}
\end{equation}%
when its initial magnitude is sufficiently large,%
\begin{equation}
\frac{\left\vert qE\right\vert }{k_{0}^{2}}\gg K_{f},\;K_{f}\gg \max \left( 
\frac{\omega _{0}}{k_{0}},1\right) \,,  \label{v}
\end{equation}%
where $K_{f}$ is a given number. We stress that condition (\ref{v})
corresponds to the most interesting case of a strong electric field where a
perturbative consideration is not applicable.

In this case, using the asymptotics of the confluent hypergeometric
functions \cite{Ederlyivol2}, we first find from expression (\ref{v13}) that
the differential mean numbers of created fermions are:%
\begin{equation}
N_{\mathbf{p},\sigma }\simeq e^{-\frac{\pi }{k_{0}}\left( \omega
_{0}-p_{x}^{\prime }\right) }\frac{\sinh \left[ \pi \left( \omega
_{0}+p_{x}^{\prime }\right) /k_{0}\right] }{\sinh \left( 2\pi \omega
_{0}/k_{0}\right) }.  \label{v16}
\end{equation}%
We note the this case is not analyzed in \cite{BRW11}, the only case when $%
\left\vert qE\right\vert \rightarrow 0$ is considered there.{\Large \ }Under
the same condition, the differential mean numbers of created scalar bosons
follow from eq. (\ref{v13b}), they are%
\begin{equation}
N_{\mathbf{p}}\simeq e^{-\frac{\pi }{k_{0}}\left( \omega _{0}-p_{x}^{\prime
}\right) }\frac{\cosh \left[ \pi \left( \omega _{0}+p_{x}^{\prime }\right)
/k_{0}\right] }{\sinh \left( 2\pi \omega _{0}/k_{0}\right) }.  \label{v16b}
\end{equation}%
Note that if the kinetic energy of final particles is big enough $\frac{%
\left\vert qE\right\vert }{k_{0}\omega _{0}}\ll 1$, the problem can be
considered perturbatively. In this case the weak time-dependent external
field violates the vacuum very small, and the corresponding pair creation
can be neglected in comparison with the main contribution given by Eqs.~(\ref%
{v16}) and (\ref{v16b}), which is formed in the momentum range (\ref{v}).

The difference in distributions (\ref{v16}) and (\ref{v16b}) that is
stipulated by the statistics is maximal for the fast varying field when $%
\omega _{0}/k_{0}\ll 1$. Then%
\begin{equation}
N_{\mathbf{p},\sigma }\simeq \frac{1}{2}\left( 1+\frac{p_{x}^{\prime }}{%
\omega _{0}}\right) ,\;N_{\mathbf{p}}\simeq \frac{k_{0}}{2\pi \omega _{0}}.
\label{v17}
\end{equation}%
In the spinless case,\emph{\ }the mean numbers $N_{\mathbf{p}}$\ given by
Eq.~(\ref{v17}) are unlimited growing.\emph{\ }This is an indication of a
big backreaction effect. Thus, we can suppose that for scalar QED the
concept of the external field is limited by the condition $2\pi
m/k_{0}\gtrsim 1$. At the same time, in the case of spinor QED, the mean
number $N_{\mathbf{p},\sigma }$, given by Eq.~(\ref{v17}), are limited $N_{%
\mathbf{p},\sigma }\leq 1$. This allows us to study fermion creation for all
possible parameters given by Eq.~(\ref{v}), using the external field concept.

It follows from Eqs.~(\ref{v16}) and (\ref{v16b}) that for large negative
longitudinal momentum,%
\begin{equation}
p_{x}^{\prime }<0,\;\left\vert p_{x}^{\prime }\right\vert /k_{0}>K_{x},
\label{cond1}
\end{equation}%
where $K_{x}\gg 1$ is a given number, the mean number of created boson and
fermion pairs is exponentially small.

In what follows we show that the main contribution to the total number of
created fermions is due to the sufficiently large positive longitudinal
momenta $p_{x}^{\prime }$ from the range%
\begin{equation}
p_{x}^{\prime }/k_{0}>K_{x},  \label{cond2}
\end{equation}%
where it is assumed that $K_{f}\gtrsim K_{x}$. In this range, it follows
from Eqs.~(\ref{v16}) and (\ref{v16b}) that%
\begin{equation}
N_{\mathbf{p},\sigma }=N_{\mathbf{p}}^{\mathrm{as}},\;N_{\mathbf{p}}^{%
\mathrm{as}}\simeq e^{-\frac{2\pi }{k_{0}}\left( \omega _{0}-p_{x}^{\prime
}\right) }  \label{v18}
\end{equation}%
both for fermions and bosons, taking into account that for bosons $N_{%
\mathbf{p}}=N_{\mathbf{p}}^{\mathrm{as}}$. We see that $N_{\mathbf{p}}^{%
\mathrm{as}}\leq 1$. Note that eq. (\ref{v18}) holds true for any
transversal energy $\sqrt{m^{2}+\mathbf{p}_{\perp }^{2}}$. In particular, if 
$\left( p_{x}^{\prime }\right) ^{2}\gg m^{2}+\mathbf{p}_{\perp }^{2}$,
distribution (\ref{v18}) can be approximated as%
\begin{equation}
N_{\mathbf{p}}^{\mathrm{as}}\simeq \exp \left( -\pi \frac{m^{2}+\mathbf{p}%
_{\perp }^{2}}{k_{0}p_{x}^{\prime }}\right) ,\,  \label{v20.2}
\end{equation}%
such that $N_{\mathbf{p}}^{\mathrm{as}}\rightarrow 1$ as $p_{x}^{\prime
}/k_{0}\rightarrow \infty $. If $\left( p_{x}^{\prime }\right) ^{2}\lesssim
m^{2}+\mathbf{p}_{\perp }^{2}$ then distribution (\ref{v18}) can be
approximated as%
\begin{equation}
N_{\mathbf{p}}^{\mathrm{as}}\lesssim \exp \left[ -\frac{2\pi }{k_{0}}\left( 
\sqrt{2}-1\right) \sqrt{m^{2}+\mathbf{p}_{\perp }^{2}}\right] .  \label{v19}
\end{equation}%
We see that this expression is exponentially small in momentum range $\sqrt{%
m^{2}+\mathbf{p}_{\perp }^{2}}/k_{0}\gtrsim p_{x}^{\prime }/k_{0}>K_{x}$.

The above analysis shows that maximum contribution to the differential
number of created fermions is provided by large positive longitudinal
momenta $p_{x}^{\prime }$, given by expression (\ref{v20.2}), with
relatively small transversal momentum $\left\vert \mathbf{p}_{\perp
}\right\vert $. Thus, taking the inequality (\ref{v}) into account, we can
conclude that the essential contribution to the total number of created
fermions is due to the longitudinal momenta $p_{x}^{\prime }$ from the wide
uniform range%
\begin{equation}
K_{x}<p_{x}^{\prime }/k_{0}<\frac{\left\vert qE\right\vert }{k_{0}^{2}}%
-K_{f},  \label{cond3}
\end{equation}%
where%
\begin{equation}
\left\vert qE\right\vert /k_{0}^{2}\gg K_{x},\;p_{x}^{\prime }\gg m.
\label{cond4}
\end{equation}

It should be noted that the contribution to the total number of created
particles from the relatively narrow momentum range of the width $K_{x}$ is
finite and of the order $K_{x}$ if $N_{\mathbf{p}}\lesssim 1$. For example,
we can use this estimation for the total number of created fermions in the
finite range of $p_{x}^{\prime }$ that is restricted by the inequality%
\begin{equation}
-K_{x}<p_{x}^{\prime }/k_{0}<K_{x}.  \label{cond5}
\end{equation}%
This contribution is much less than the contribution from a very wide range (%
\ref{cond3}). The same is true for bosons when $2\pi m/k_{0}\gtrsim 1$.\
That is why the contribution to the total number of created fermions in the
range (\ref{cond3}) is the main contribution. The main contribution to the
total number of created bosons is due to the range (\ref{cond3}) for the
slowly decaying electric field when $2\pi m/k_{0}\gtrsim 1$.

Note that if $\left( \omega _{0}-p_{x}^{\prime }\right) /k_{0}\gg 1,$ WKB
approximation holds true for $N_{\mathbf{p},\sigma }$ given by Eq.~(\ref{v18}%
) . In this domain expression (\ref{v18}) coincides exactly with an
estimation, obtained previously in \cite{Spokoinyi} using the semiclassical
consideration, while our approximation (\ref{v18}) is valid for any value of 
$\left( \omega _{0}-p_{x}^{\prime }\right) /k_{0}$ and our exact results,
given by Eqs.~~(\ref{v16}) and (\ref{v16b}), are quite different from the
semiclassical ones.

\subsubsection{Total quantities}

The obtained distribution $N_{n}=\left\vert g(_{-}|^{+})\right\vert ^{2}$
plays the role of a cut-off factor in the integral over momenta (\ref{e1b})
for the total number of created particles (there for bosons $J_{(d)}=1)$.
However, for bosons, this result is valid only if the electric field decays
slowly enough, $2\pi m/k_{0}\gtrsim 1$. Then the total number of created
particles can be represented by its main contribution in the range (\ref%
{cond3}) as follows 
\begin{equation}
N\approx \frac{VJ_{(d)}}{(2\pi )^{d-1}}\int_{p_{x}^{\min }}^{p_{x}^{\max
}}dp_{x}^{\prime }N_{p_{x}}^{\mathrm{as}},\ \ N_{p_{x}}^{\mathrm{as}}=\int d%
\mathbf{p}_{\bot }N_{\mathbf{p}}^{\mathrm{as}},  \label{t1}
\end{equation}%
where $N_{\mathbf{p}}^{\mathrm{as}}$ is given by eq. (\ref{v20.2}) and 
\begin{equation}
p_{x}^{\max }=\frac{\left\vert qE\right\vert }{k_{0}}-K_{f}k_{0},\ \
p_{x}^{\min }=K_{x}k_{0}.  \label{t2}
\end{equation}%
Integrating over $\mathbf{p}_{\bot }$ and taking into account that $%
p_{x}^{\prime }\gg m$, we find that the total number of created particles
with a given longitudinal momentum reads

\begin{equation}
N_{p_{x}}^{\mathrm{as}}\approx \left( k_{0}p_{x}^{\prime }\right)
^{d/2-1}\exp \left( -\frac{\pi m^{2}}{k_{0}p_{x}^{\prime }}\right) .
\label{t4}
\end{equation}

Using Eq.~(\ref{t4}), we represent the integral (\ref{t1}) in the form%
\begin{equation}
N\approx \frac{VJ_{(d)}}{(2\pi )^{d-1}}\left( k_{0}\right) ^{d/2-1}Y^{\left(
1\right) },  \label{t5}
\end{equation}%
where $Y^{\left( 1\right) }$ is the particular case of the integral%
\begin{equation}
Y^{\left( k\right) }=\int_{p_{x}^{\min }}^{p_{x}^{\max }}dp_{x}^{\prime
}\left( p_{x}^{\prime }\right) ^{d/2-1}\exp \left( -k\frac{\pi m^{2}}{%
k_{0}p_{x}^{\prime }}\right) ,\ \ k=1,2,\ldots \ .  \label{t6}
\end{equation}%
Taking into account that $\left\vert qE\right\vert /k_{0}^{2}\gg
K_{f}\gtrsim K_{x}$, we obtain that the integral (\ref{t6}) is independent
on the given numbers $K_{f}$ and $K_{x}$ in the leading order term
approximation. If $m\neq 0$ then the integral (\ref{t6}) in this
approximation can be expressed via the incomplete gamma function as%
\begin{equation}
Y^{\left( k\right) }\approx \left( \frac{k_{0}}{\pi m^{2}k}\right)
^{d/2}\Gamma \left( -\frac{d}{2},k\frac{\pi m^{2}}{\left\vert qE\right\vert }%
\right) ,\ \ k=1,2,\ldots \ \ .  \label{t7}
\end{equation}

Note that the representation (\ref{t7}) is suitable when the electric field
is weak enough, $k\pi m^{2}/\left\vert qE\right\vert \gg 1$. In this case
one can use the following asymptotics of the incomplete gamma function,%
\begin{equation}
\Gamma \left( -\frac{d}{2},k\frac{\pi m^{2}}{\left\vert qE\right\vert }%
\right) \approx \exp \left( -k\frac{\pi m^{2}}{\left\vert qE\right\vert }%
\right) \left( k\frac{\pi m^{2}}{\left\vert qE\right\vert }\right) ^{-d/2-1}.
\label{t8}
\end{equation}

For the case of a strong field, when $k\pi m^{2}/\left\vert qE\right\vert
\ll 1$, where the case of massless fermions is included too, we find in the
leading order term approximation that%
\begin{equation}
Y^{\left( k\right) }\approx \frac{2}{d}\left( \frac{\left\vert qE\right\vert 
}{k_{0}}\right) ^{d/2}.  \label{t9}
\end{equation}%
Then the total number of particles created from vacuum is%
\begin{equation}
N^{\mathrm{strong}}\approx \frac{VJ_{(d)}}{(2\pi )^{d-1}}\frac{2\left(
\left\vert qE\right\vert \right) ^{d/2}}{k_{0}d}.  \label{t10}
\end{equation}

Finally taking into account the above results, we can represent the
probability of a vacuum to remain a vacuum, defined by Eq.(\ref{vacprob}), as%
\begin{equation}
P_{v}\approx \exp \left\{ -\frac{VJ_{(d)}}{(2\pi )^{d-1}}\sum_{k=0}^{\infty }%
\frac{(-1)^{(1-\kappa )k/2}}{(k+1)^{d/2}}\left( k_{0}\right)
^{d/2-1}Y^{\left( k+1\right) }\right\} ,  \label{t11}
\end{equation}%
where $Y^{\left( k+1\right) }$ is given by the integral (\ref{t6}) and can
be represented in the leading term approximation with the help of Eqs.~(\ref%
{t7}) and (\ref{t9}), respectively. For the strong field case we find that
the probability $P_{v}$ is determined by the total number of created
particles 
\begin{equation}
P_{v}^{\mathrm{strong}}=\exp \left\{ -\mu N^{\mathrm{strong}}\right\} ,\;\mu
=-\sum_{k=0}^{\infty }\frac{(-1)^{(1-\kappa )k/2}}{(k+1)^{d/2}}.  \label{t12}
\end{equation}

One can see that dependence of the total number of particles created from
vacuum by the strong exponential field on the field magnitude and space-time
dimensions mimics the case of particle creation by strong $T$-constant
electric field $E$ (see \cite{GavGit96}) for big $T$ and with the
identification $T=2\left( k_{0}d\right) ^{-1}$. It is due to the effect of
saturation for the distribution $N_{n}\rightarrow 1$ in the wide uniform
range of initial longitudinal momentum, where there is a big increment of
the kinetic momentum, $\left\vert qE\right\vert /k_{0}$, and $\left\vert
qE\right\vert T$, for both cases, respectively.

Let us consider two strong $T$-exponential electric fields of the same
magnitude $E$ but with distinct parameters $k_{0}^{(I)}$ and $%
k_{0}^{(II)}\ll k_{0}^{(I)}$. Let they create from a vacuum the total
numbers of particles $N^{(I)}$ and $N^{(II)}$, respectively. One can see
from Eq.~(\ref{t10}) that $N^{(II)}\gg N^{(I)}$, that is, the electric field
of more long effective duration creates much more pairs. The total number of 
$\mathrm{out}$-particles created from $\mathrm{in}$-vacuum due to a
decreasing exponential field is the same with the total number of particles
created from a vacuum due to a increasing exponential field provided that
the modulus of potential difference is the same for both cases. That is, we
can consider $N^{(I)}$ as the total number of particles created from a
vacuum due to the increasing field. We see that if $k_{0}^{(II)}\ll
k_{0}^{(I)},$ the main contribution to particle creation by external
electric field that switches on fast enough and then slowly decreases is due
to its decreasing part, whereas the contribution from the increasing part of
the field is relatively small. In particular, the exponentially decaying
electric field can be treated as one, which is created by an external
current that switches on fast enough and then is slowly switching off
because of some dissipation processes. Thus we see that the exponentially
decaying electric field under consideration allows one to study the role of
switching on and switching off processes.

\section*{Acknowledgements}

TCA acknowledges the support of FAPESP under the contract 2013/00840-9. SPG
thanks FAPESP for a support and University of S\~{a}o Paulo for the
hospitality. DMG is grateful to the Brazilian foundations FAPESP and CNPq
for permanent support. The work of SPG and DMG is also partially supported
by the Tomsk State University Competitiveness Improvement Program.

\appendix

\section{Pair creation in a homogeneous electric field\label{App}}

Following general consideration in \cite{GavGit96}, we recall in this
Appendix some basic elements of the generalized Furry representation \cite%
{Gitman77,FGS,GavGT06} that is used to describe vacuum instability in a
strong external time-dependent electric field.

For the particular case of a homogeneous electric field, we assume that the
potential $A_{1}(t)$ ($A_{\mu }(t)=0,\ \mu \neq 1$), is constant for $%
t<t_{1} $ and for $t>t_{2}$. Therefore, the initial (at $t<t_{1}$) and the
final (at $t>t_{2}$) vacua are vacuum states of $\mathrm{in}$- and $\mathrm{%
out}$- free particles which correspond to the constant effective potentials $%
A_{1}\left( t_{1}\right) $ and $A_{1}\left( t_{2}\right) $, respectively.
During the time interval $t_{2}$ $-t_{1}$ $=T$, the quantum Dirac field
interacts with the time-dependent effective potential $A_{1}\left( t\right) $%
. In the general case, the initial and final vacua are different. We
introduce an initial set of creation and annihilation operators $%
a_{n}^{\dagger }($\textrm{in}$)$, $a_{n}($\textrm{in}$)$ of \textrm{in}%
-particles (electrons), and operators $b_{n}^{\dagger }($\textrm{in}$)$, $%
b_{n}($\textrm{in}$)$ of \textrm{in}-antiparticles (positrons), the
corresponding \textrm{in}-vacuum, being $|0,$\textrm{in}$\rangle $, and a
final set of creation and annihilation operators $a_{n}^{\dagger }($\textrm{%
out}$)$, $a_{n}($\textrm{out}$)$ of \textrm{out}-electrons and operators $%
b_{n}^{\dagger }($\textrm{out}$)$, $b_{n}($\textrm{out}$)$ of \textrm{out}%
-positrons, the corresponding \textrm{out}-vacuum, being $|0,$\textrm{out}$%
\rangle $. Thus for any quantum number $n$, we have%
\begin{align}
& a_{n}(\mathrm{in})|0,\mathrm{in}\rangle =b_{n}(\mathrm{in})|0,\mathrm{in}%
\rangle =0,  \notag \\
& a_{n}(\mathrm{out})|0,\mathrm{out}\rangle =b_{n}(\mathrm{out})|0,\mathrm{%
out}\rangle =0.  \label{3.5}
\end{align}%
In both cases, by $n=\left( \mathbf{p},\sigma \right) $ we denote complete
sets of quantum numbers that describe both $\mathrm{in}$- and $\mathrm{out}$%
- particles and antiparticles. The $\mathrm{in}$-operators and the \textrm{%
out}-operators obey the canonical anticommutation relations. The above $%
\mathrm{in}$- and $\mathrm{out}$-operators are defined by two decompositions
of the quantum Dirac field $\Psi (x)$ in the exact solutions of the Dirac
equation, 
\begin{align}
\Psi \left( x\right) & =\sum_{n}\left[ a_{n}\left( \mathrm{in}\right) \left.
_{+}\!\psi _{n}\left( x\right) \right. +b_{n}^{\dagger }\left( \mathrm{in}%
\right) \left. _{-}\!\psi _{n}\left( x\right) \right. \right] \,  \notag \\
& =\sum_{n}\left[ a_{n}\left( \mathrm{out}\right) \left. ^{+}\!\psi
_{n}\left( x\right) \right. +b_{n}^{\dagger }\left( \mathrm{out}\right)
\left. ^{-}\!\psi _{n}\left( x\right) \right. \right] \,.  \label{fields}
\end{align}%
Thus, the $\mathrm{in}$-operators are associated with a complete orthonormal
set of solutions $\left\{ _{\zeta }\psi _{n}(x)\right\} $ (we call it the $%
\mathrm{in}$-set) of Eq.~(\ref{iint0}), where $\zeta =+$ stays for electrons
and $\zeta =-$ for positrons. Their asymptotics at $t<t_{1}$ are wave
functions of free particles in the presence of a constant electric potential 
$A_{1}\left( t_{1}\right) $. The $\mathrm{out}$-operators are associated
with another complete orthonormal $\mathrm{out}$-set of solutions $\left\{
^{\zeta }\psi _{n}\left( x\right) \right\} $ of Eq.~(\ref{iint0}). Their
asymptotics at $t>t_{2}$ are wave functions of free particles in the
presence of a constant electric potential $A_{1}\left( t_{2}\right) $.

The inner product between two solutions $\psi \left( x\right) $ and $\psi
^{\prime }\left( x\right) $ of the Dirac equation on $t$-\textrm{const}
hyperplane,%
\begin{equation}
\left( \psi ,\psi ^{\prime }\right) =\int \psi ^{\dag }\left( x\right) \psi
^{\prime }\left( x\right) d\mathbf{x},  \label{IP}
\end{equation}%
is time-independent. Then, taking into account the structure (\ref{t4.10})
and initial or final forms of the functions $_{\zeta }\psi _{n}(x)$ and $%
^{\zeta }\!\psi _{n}$, respectively, one finds the orthonormality relations:%
\begin{equation}
\left( \left. _{\zeta \!}\psi _{n}\right. ,\left. _{\zeta ^{\prime }}\!\psi
_{n^{\prime }}^{\prime }\right. \right) =\delta _{n,n^{\prime }}\delta
_{\zeta ,\zeta ^{\prime }}\,,\ \ \left( \left. ^{\zeta }\!\psi _{n}\right.
,\left. ^{\zeta ^{\prime }}\!\psi _{n^{\prime }}^{\prime }\right. \right)
=\delta _{nn^{\prime }}\delta _{\zeta ,\zeta ^{\prime }}\,.  \label{t4.34}
\end{equation}%
Here we apply the standard QFT volume regularization assuming that all the
processes are confined in a big $D$ dimensional space box with the volume $V$%
. $\mathrm{In}$- and $\mathrm{out}$-solutions with given quantum numbers $n$
are related by linear transformations of the form 
\begin{align}
^{\zeta }\psi _{n}\left( x\right) & =g(_{+}\mid ^{\zeta })\,_{+}\psi
_{n}\left( x\right) +g(_{-}\mid ^{\zeta })\,_{-}\psi _{n}\left( x\right) \,,
\notag \\
_{\zeta }\psi _{n}\left( x\right) & =g\left( ^{+}|_{\zeta }\right)
\,^{+}\psi _{n}\left( x\right) +g\left( ^{-}|_{\zeta }\right) \,^{-}\psi
_{n}\left( x\right) ,  \label{3.12}
\end{align}%
where the coefficients $g$ are defined via the inner products of these sets, 
\begin{equation}
\left( _{\zeta }\psi _{n},^{\zeta ^{\prime }}\psi _{n`}\right) =\delta
_{n,n^{\prime }}g\left( {}_{\zeta }|{}^{\zeta ^{\prime }}\right) ,\ \
g\left( ^{\zeta ^{\prime }}|_{\zeta }\right) =g\left( _{\zeta }|^{\zeta
^{\prime }}\right) ^{\ast }.  \label{3.13}
\end{equation}%
These coefficients satisfy the unitarity relations, which follow from the
orthonormality relations (\ref{t4.34}), and can be expressed in terms of two
of them, e.g., of $g\left( _{+}\left\vert ^{+}\right. \right) $ and $g\left(
_{-}\left\vert ^{+}\right. \right) $. However, even these coefficients are
not completely independent, 
\begin{equation}
\left\vert g\left( _{-}\left\vert ^{+}\right. \right) \right\vert
^{2}+\left\vert g\left( _{+}\left\vert ^{+}\right. \right) \right\vert
^{2}=1.  \label{3.15}
\end{equation}

A linear canonical transformation (Bogolyubov transformation) between $%
\mathrm{in}$- and $\mathrm{out}$- operators which can be derived from Eq.~(%
\ref{fields}) has the form 
\begin{align}
a_{n}\left( \mathrm{out}\right) & =g\left( ^{+}|_{+}\right) a_{n}(\mathrm{in}%
)+g\left( ^{+}|_{-}\right) b_{n}^{\dagger}(\mathrm{in}),  \notag \\
b_{n}^{\dagger}\left( \mathrm{out}\right) & =g\left( ^{-}|_{+}\right) a_{n}(%
\mathrm{in})+g\left( ^{-}|_{-}\right) b_{n}^{\dagger}(\mathrm{in}).
\label{canonical}
\end{align}
Then one can see that all the information about electron-positron creation,
annihilation, and scattering in an external field can be extracted from the
coefficients $g\left( {}_{\zeta}|{}^{\zeta^{\prime}}\right) $.

One of the most important quantity for the study of particle creation is
differential mean number of created particles, defined as the expectation
value of $\mathrm{out}$ number operator with respect to the $\mathrm{in}$%
-vacuum,%
\begin{equation}
N_{n}=\langle 0,\mathrm{in}|a_{n}^{\dagger }(\mathrm{out})a_{n}(\mathrm{out}%
)|0,\mathrm{in}\rangle =|g(_{-}|^{+})|^{2}\,.  \label{difnumber}
\end{equation}%
It is equal to the mean number of particle-antiparticle pairs created. The
total number of created particles is obtained by the summation over the
quantum numbers $n$,%
\begin{equation}
N=\sum_{n}N_{n}\,.  \label{totnumber}
\end{equation}%
The probability of a vacuum to remain a vacuum, is defined as%
\begin{equation}
P_{v}=\left\vert \langle 0,\mathrm{out}|\left. 0,\mathrm{in}\right\rangle
\right\vert ^{2}=\exp \left\{ \kappa \sum_{n}\ln \left( 1-\kappa
N_{n}\right) \right\} \,,  \label{vacprob}
\end{equation}%
where $\kappa =+1$ for fermions and $\kappa =-1$ for bosons. The probability
of the electron scattering $P(+|+)_{n,n^{\prime }}$ and the probability of a
pair creation $P(-+|0)_{n,n^{\prime }}$ are, respectively 
\begin{align}
& P(+|+)_{n,n^{\prime }}=|\langle 0,\mathrm{out}|a_{n}(\mathrm{out}%
)a_{n^{\prime }}^{\dagger }(\mathrm{in})|0,\mathrm{in}\rangle |^{2}=\delta
_{n,n^{\prime }}\frac{1}{1-\kappa N_{n}}P_{v}\;,  \notag \\
& P(-+|0)_{n,n^{\prime }}=|\langle 0,\mathrm{out}|b_{n}(\mathrm{out}%
)a_{n^{\prime }}(\mathrm{out})|0,\mathrm{in}\rangle |^{2}=\delta
_{n,n^{\prime }}\frac{N_{n}}{1-\kappa N_{n}}P_{v}\;.  \label{3.18}
\end{align}%
The probabilities for a positron scattering and a pair annihilation are
given by the same expressions $P(+|+)$ and $P(-+|0)$, respectively.


\begin{thebibliography}{99}
\bibitem{Sch51} J. Schwinger, Phys. Rev. \textbf{82}, 664 (1951).

\bibitem{Nik70} A. I. Nikishov, Zh. Eksp. Teor. Fiz. \textbf{57}, 1210
(1969) [Transl. Sov. Phys. JETP \textbf{30}, 660 (1970)]; A. I. Nikishov, in 
\emph{Quantum Electrodynamics of Phenomena in Intense Fields}, Proc. P.N.
Lebedev Phys. Inst. \textbf{111}, 153 (Nauka, Moscow 1979).

\bibitem{Gitman77} D. M. Gitman, J. Phys. A: Math. Gen. \textbf{10}, 2007
(1977); E. S. Fradkin and D. M. Gitman, Fortschr. Phys. \textbf{29,} 381
(1981)

\bibitem{FGS} E.S. Fradkin, D.M. Gitman and S.M. Shvartsman, \emph{Quantum
Electrodynamics with Unstable Vacuum} (Springer-Verlag, Berlin, 1991).

\bibitem{GMR85} W. Greiner, B. M\"{u}ller and J. Rafelsky, \emph{Quantum
electrodynamics of strong fields} (Springer-Verlag, Berlin, 1985).

\bibitem{BirDav82} N.~D.~Birrell and P.~C.~W.~Davies, \emph{Quantum Fields
in Curved Space} (Cambridge University Press, Cambridge, 1982); A.~A.~Grib,
S.~G.~Mamaev, and V.~M.~Mostepanenko, \emph{Vacuum Quantum Effects in Strong
Fields} (Friedmann Laboratory Publishing, St. Petersburg, 1994); R.~Ruffini,
G.~Vereshchagin and S.~Xue, Phys. Rep. \textbf{487}, 1 (2010).

\bibitem{GavGT06} S.~P.~Gavrilov, D.~M.~Gitman, and J.~L.~Tomazelli, Nucl.
Phys. B \textbf{795, }645 (2008) [hep-th/0612064].

\bibitem{Dun09} G.~V.~Dunne, Eur. Phys. J. D \textbf{55}, 327 (2009).

\bibitem{castroneto} A. H. Castro Neto \emph{et al}, Rev. Mod. Phys. \textbf{%
81,} 109 (2009); N. M. R. Peres, Rev. Mod. Phys. \textbf{82,} 2673 (2010);
M.A.H. Vozmediano, M.I. Katsnelson, F. Guinea, Phys. Rep. \textbf{496}, 109
(2010).

\bibitem{dassarma} D. Das Sarma, S. Adam, E. H. Hwang and E. Rossi, Rev.
Mod. Phys. \textbf{83,} 407 (2011).

\bibitem{GavGitY12} S. P. Gavrilov, D. M. Gitman and N. Yokomizo, Phys. Rev.
D. \textbf{86}, 125022 (2012) [arXiv:1207.1749].

\bibitem{DumDun13} C.K. Dumlu, G.V. Dunne, Phys. Rev. D \textbf{84}, 125023
(2011) [arXiv:1110.1657].

\bibitem{DS05} G.V. Dunne, C. Schubert, Phys.Rev. D \textbf{72} (2005)
105004 [arXiv:hep-th/0507174]; G.V. Dunne, H. Gies, C. Schubert, Q.-h. Wang,
Phys. Rev. D \textbf{73}, 065028, (2006) [arXiv:hep-th/0602176].

\bibitem{I14} A. Ilderton, JHEP \textbf{09} (2014) 166 [arXiv:1406.1513 ];
C. Schneider, R. Sch\"{u}tzhold, \emph{Dynamically assisted Sauter-Schwinger
effect in inhomogeneous electric fields}, arXiv:1407.3584.

\bibitem{NarNik70} N.B. Narozhny and A.I. Nikishov, Sov. J. Nucl. Phys.
(USA) \textbf{11}, 596 (1970).

\bibitem{DunHal98} G. Dunne and T. Hall, Phys. Rev. D \textbf{58}, 105022
(1998).

\bibitem{BagGiS75} V. G. Bagrov, D. M. Gitman, and Sh. M. Shvartsman, Zh.
Eksp. Teor. Fiz. \textbf{68}, 392 (1975) (Sov. Phys. JETP \textbf{41}, 191
(1975)).

\bibitem{GavGit96} S. P. Gavrilov and D. M. Gitman, Phys. Rev. D \textbf{53}%
, 7162 (1996) [hep-th/9603152].

\bibitem{GavGit08} S.P. Gavrilov and D.M. Gitman, Phys. Rev. Lett. \textbf{%
101}, 130403 (2008) [arXiv:0805.2391]; S. P. Gavrilov and D. M. Gitman,
Phys. Rev. D \textbf{78}, 045017 (2008) [arXiv:0709.1828].

\bibitem{NarNIk74} N.B. Narozhny and A.I. Nikishov, Sov. Phys. JETP \textbf{%
38}, 427 (1974); V.M. Mostepanenko and V.M. Frolov, Sov. J. Nucl. Phys.
(USA) \textbf{19}, 451 (1974).

\bibitem{TTW01} T. N. Tomaras, N. C. Tsamis, R. P. Woodard, JHEP\textbf{\
0111} (2001) 008 [arXiv:hep-th/0108090].

\bibitem{BagGi90} V.G. Bagrov and D.M. Gitman, \emph{Exact Solutions of
Relativistic Wave Equations} (Kluwer, Dordrecht 1990); \emph{Dirac Equation
and its Solutions} (de Gruyter, Boston, 2014)

\bibitem{Spokoinyi} B. L. Spokoinyi, Yad. Fiz. \textbf{36}, 474 (1982)
[English transl. Sov. J. Nucl. Phys. \textbf{36}, 277 (1982)]; Phys. Lett. A 
\textbf{88}, 328 (1982).

\bibitem{BRW11} I. Bialynicki-Birula, \L . Rudnicki, A, Wienczek, \emph{Pair
creation by time-dependent electric fields: Analytic solutions},
arXiv:1108.2615.

\bibitem{Ederlyivol2} \emph{Higher Transcendental functions} (Bateman
Manuscript Project), edited by A. Erdelyi \emph{et al.} (McGraw-Hill, New
York, 1953), Vols. 1.
\end{thebibliography}
\end{document}